# How a user's personality influences content engagement in social media

Nathan O. Hodas[1], Ryan Butner[2], Court Corley[1]

[1]Pacific Northwest National Laboratory, Richland, WA
{nhodas,court}@pnnl.gov
[2]Monsanto, St. Louis, MO
ryan.scott.butner@monsanto.com

**Abstract.** Social media presents an opportunity for people to share content that they find to be significant, funny, or notable. No single piece of content will appeal to all users, but are there systematic variations between users that can help us better understand information propagation? We conducted an experiment exploring social media usage during disaster scenarios, combining electroencephalogram (EEG), personality surveys, and prompts to share social media, we show how personality not only drives willingness to engage with social media, but also helps to determine what type of content users find compelling. As expected, extroverts are more likely to share content. In contrast, one of our central results is that individuals with depressive personalities are the most likely cohort to share informative content, like news or alerts. Because personality and mood will generally be highly correlated between friends via homophily, our results may be an import factor in understanding social contagion.

## 1 Introduction

Whether for disaster response, advertising campaigns, or general entertainment, people leverage social media to spread information to wide and varied audiences. When crafting a message on social media, authors may attempt to consider humor (Evers et al. 2013), trustworthiness (Kietzmann et al. 2011), or timeliness (Lee and Ma 2012), among other factors, to increase the reach of their message. Authors may not consider the personality or mood of target users when anticipating the impact and propagation of their messages. Systematic biases in target populations will confound attempts to understand social contagion (Hodas and Lerman 2014). Because of homophily, personality types will not be randomly distributed in the social network, and users will be exposed to content biased by the personality of their friends (Hodas et al. 2013). It is important to better understand the link between personality, mood and social contagion.

In this paper, we reveal a systematic link between personality type and mood, brain response, and the type of content people choose to share online. Although it comes as no surprise that there is a relationship between how someone uses social media and their personality (Ryan and Xenos 2011; Correa et al. 2010; Hughes et al. 2012), this is the first experiment that measured both the user's present mood and personality, quantitative measures of engagement and interest, as well as their final reactions to the content. We originally conducted this research in the context of understanding user's responses to natural disasters via social media. In the methods section below we explore this experiment in detail, including how users were assessed for personality and mood, were shown videos describing the disasters, then asked to share (or not) tweets and emergency alerts, all while being continuously monitored via electroencephalogram (EEG). In this way, we have quantitative measures of personality, attention, and action.

The main finding of this paper is that users systematically prefer different types of content, and that this content depends on their personality and mood in significant ways. The different types of content, such as "informative", "social", or "sympathetic," which we describe below, each resonate differently depending on personality and mood. For example, as one would expect, extroverts are more likely to share any content, consistent with previous findings. However, we also find that users that score highest on measures of depression were more likely to share informative messages, compared to the least depressed users.

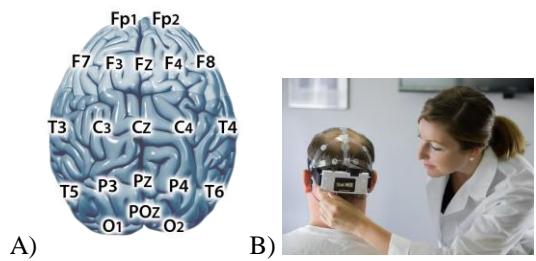

**Fig. 1.** B-Alert X24 EEG System. A) Standard 10-20 Montage B) 24 Channel Wireless Headset

Because of correlation between content, type of information, and personality, we show that different types of personalities will be more responsive to different kinds of information campaigns.

The paper is presented as follows. First, we discuss the unique experiment we conducted and describe the methods we used to understand user behavior. Next, we describe the results of our experiments and discuss their importance to understanding how personality impacts information transmission. Lastly, we compare our work to the existing literature. Our unique contribution is to separate personality from engagement using brain monitoring, revealing that the personality and mood of targeted users plays a significant role in determining the type of information that gets selected by users to share.

## 2 Methods

The purpose of this project, using an electroencephalogram (EEG) data-driven approach, was to evaluate the physiological response of individuals to social media content within the context of emergency situations. This approach allowed an analysis of how subjects perceive disaster alerts, observation of the level of attention elicited in subjects, and observation of subjects' response to the question of whether to share such alerts with their peers over a social media platform. The authors chose Twitter as the target platform because its 140-character limit is most representative of the current 90-character allowance for cell phone-based alerts about weather-related emergencies and because it is a ubiquitous platform.

The experiment evaluated test subjects' willingness to share messages to their own personal social network. These messages included Wireless Emergency Alerts (WEA) and tweets associated with five different types of disasters. Among the tweets were messages conveying sympathy for the victims of disasters, and other forms of sociable communication over the social network. We asked subjects, within the context of a natural disaster, to evaluate how important they perceived various forms of communication about several disasters (specifically, a blizzard, flood, gas leak, hurricane, and tornado). By evaluating their responses alongside their physiological response to the messages, the experiment measured their willingness to disseminate information about disasters and analyze the underlying cognitive models that drive their perceptions and reactions about different types of disasters.

The Twitter messages used in this study were a combination of real messages posted on Twitter during that disaster and disaster alerts sent by news stations and other emergency alert services within a defined geographic region surrounding the site of a declared emergency. For disasters that were declared at a definite point (such as tornados), tweets were collected from within the surrounding 25-mile radius. For disasters that affected broader swathes of land (such as hurricanes or blizzards), tweets were selected from within the entire region being alerted for a weather emergency. We collected tweets associated from the following disasters:

- blizzard, a winter storm that struck South Dakota in October 2013.
- flash flood, an episode of flooding that occurred in southern California in the summer of 2013.
- gas leak, an incident that occurred in Alamo, California on July 24, 2013.
- hurricane, across the northeastern United States, where Hurricane Sandy made landfall in late October 2012.
- tornado, a tornado that struck Moore, Oklahoma on May 20, 2013.

Experimental Data Collection
Scientists at the Advanced Brain Monitoring (ABM) laboratory in Carlsbad, CA acquired electroencephalography (EEG) data from 51 participants during an experiment to evaluate the ways in which people perceive different kinds of disasters. The ABM wireless B-Alert® EEG sensor headset, a lightweight, easy-to-apply system was used to acquire 20 channels of data from sites: Fz, Fp1, Fp2, F3, F4, F7, F8, Cz, C3, C4, Pz, P3, P4, POz, T3, T4, T5, T6, O1, and O2, all referenced to linked mastoids.

The experiment presented each subject with five disasters in randomized order. A random benchmark assessment of six neutral tweets (i.e., tweets that were not related to any disaster) was presented either immediately before or after the set of disaster blocks (i.e., first or last). Immediately before each disaster block, subjects were first shown a 5-minute newsreel video depicting news coverage for the disaster type they were about to evaluate.

After the newsreel ended, subjects were presented with a series of 50 WEA and Twitter messages for each of the five types of disasters. The testbed presented each message for a minimum of six seconds before the user was permitted to answer, to give the user time to read the message and reduce impulsive responses. The testbed then asked participants if they would share the message on social media. The subjects were required to use the keyboard to respond

"yes" or "no" before moving on to the next message. Each subject received the disasters and associated messages in random order.

## 2.1 Description of the EEG Data

EEG data was time-locked to both the onset of each stimulus (messages or videos) and to each response (yes/no). All EEG data were acquired at a 256Hz sampling rate (i.e., there are 256 measurements of brain activity taken every second) to provide a high level of fidelity in the analysis. The analysis of this data focused on measuring the brain's electrical response resulting from exposure to a particular cognitive or sensory event. ABM filtered the signal to remove blinks and other known signal confounders, and the remaining raw signal was used for analysis.

ABM measured each participant's head to ensure proper sizing and positioning of the 20-channel sensor cap, shown in Fig. 1. ABM designed the headset to position sensors over all cortical regions in accordance with the International 10-20 system, also shown in Fig. 1. While the association between a specific location on the human scalp and the precise activity occurring in the brain beneath it is not an exact correlation in all cases, in practice, EEG activity recorded over specific regions has been associated with particular functions or responses in subjects.

## 2.2 Personality assessment

Each user completed multiple surveys designed to assess their personality and mood. Here, we report on two of those surveys. This includes the NEO personality inventory to assess the "Big 5" personality traits: openness, conscientiousness, extraversion, agreeableness and neuroticism. To assess current, transient mood, the subjects completed the Profile of Mood States (POMS). The mood states include anxiety, depression, anger, fatigue, vigor, and confusion.

## 2.3 Content Annotation

Each message, either an emergency alert or a tweet, was hand annotated by subject matter experts to be in one of three categories: informative, social, or sympathetic. Although these categories were chosen to be particularly apropos for disaster scenarios, they may be generalized to other domains as follows:

*Informative* – These messages contain objective information related to an event intended for a user to factor into their decision-making. Examples include:

- ```
  12" of snow so far just NW of Rapid
  City, SD. Sustained winds over 40
  mph. Blizzard warning until tomorrow
  morning.
  ```
- ```
  Flash flood warning #palmsprings
  #coachellavalley @[username]
  ```
- ```
  Superstorm Sandy will hit east coast
  USA - 140 km/hour winds Monday _
  Connecticut, New York, New Jersey
  #amsterdam #haarlem #rotterdam
  ```
- ```
  Tornado emergency for Moore #OK from
  @[username] Take shelter now
  ```
- ```
  Still a tornado warning for Paul's
  Valley area. Continue to be taking
  cover.
  ```

*Social* – These messages convey information about an event but in a way that emphasizes the social aspects of the event or is used as a means to communicate informal information about the event. Examples include:

- ```
  @[username] we had the worst bliz-
  zard in the history of souf dadoka
  ```
- ```
  And we have power! 26 hrs w/o elec-
  tricity and heat is #funtimes
  #BlackHills #blizzard
  ```
- ```
  Blizzard is going on and also light-
  ing and thunder
  ```
- ```
  Blizzard still raging on. No power
  for almost two hours. Hello October.
  ```
- ```
  Another flash flood warning? Uh ohh
  I hope there isn't any more thunder
  storms
  ```

*Sympathetic* – These messages convey explicit emotions or sympathy specifically related to others involved in the event. Examples include:

- ```
  @[username]: South Dakota's state
  veterinarian believes up to 20,000
  cattle died in a blizzard.  sadly,
  we made cnn
  ```
- ```
  @[username]: West river South Dakota
  cattle losses may total 25% of herd.
  25% of a herd of +2,000,000 head.
  #blizzard2013â€• sad sad deal here
  ```
- ```
  a mother was trying to drive her 2
  young sons to Brooklyn because she
  was scared about the storm& a huge
  wave hit them&two baby boys gone
  ```
- ```
  Death toll now up to 96 from #Sandy.
  #RIP to the beautiful souls.
  ```

- Please pray for all those in Moore Oklahoma #tornados have devastated the area.

## 3  Results and Discussion

The results of this study reveal that the personality types of social media users impact their preferences or willingness to share certain forms of content. During the experiment, users exhibited distinct preferences for specific types of content that corresponded to their scores on various personality dimensions. These preferences became most apparent when cohorts of subjects with the highest scores for specific personality dimensions compared against cohorts of subjects with the lowest scores. These preferences were observed both in the levels of brain activity when viewing the disaster context videos, the frequency at which the subjects shared content, and in their corresponding EEG signatures when choosing to share content.

The frontal regions of the brain (designated by sites named with an "F") were of particular interest in this study, as this region is most strongly associated with executive function and decision-making. Among the 20 channels examined in the study, the subject response to the disaster context videos and the brain activity during decision-making for the individual tweets were observed to be most prominent in the F7, F8, Fp1, and Fp2 channels (all odd number regions are left hemisphere and even numbers are right hemisphere). Channels associated with other brain regions did not exhibit any noteworthy response to either the context videos or messages.

Specifically, subjects exhibited the higher levels of brain activity over the Fp1 and Fp2 regions channels during presentation of the context videos. Subjects similarly exhibited greater levels of activity over the F7 and F8 regions when determining if they would share specific messages with their social network. The left and right frontal regions appear to drive the decision making process to share specific messages, a finding consistent with prior evidence linking these regions to motivation and mood regulation (Davidson 2004). Conversely, EEG activity over the Fp1 and Fp2 regions is commonly associated with logical or emotional attention, judgment, and decision making (Chen et al. 2015).

As shown in Fig. 2, during presentation of the context videos, subjects with the highest scores for depressive, fatigued, or confused personalities from the POMS personality test battery exhibited the lower engagement EEG scores. Similarly, subjects with the most extroverted, open, and agreeable personality scores according to the NEO personality test battery exhibited stronger signs of engagement and attention than their counterparts with the lowest scores on these metrics. All differences were significant to at least $p<0.05$.

Because the activity during the video is indicative of attention and engagement in the task—users whose mood is characterized as depressed, fatigued or confused—will generally be less engaged with the content on social media, even during controlled conditions (Fig. 2a). Conversely, highly agreeable, extroverted or open users appear to more readily engage in the videos than their lower scoring counterparts (Fig. 2b). As an aside, empirical studies of social contagion have difficulty distinguishing if users don't spread a message because they didn't like it or if they didn't see it, i.e. low visibility (Hodas and Lerman 2014).

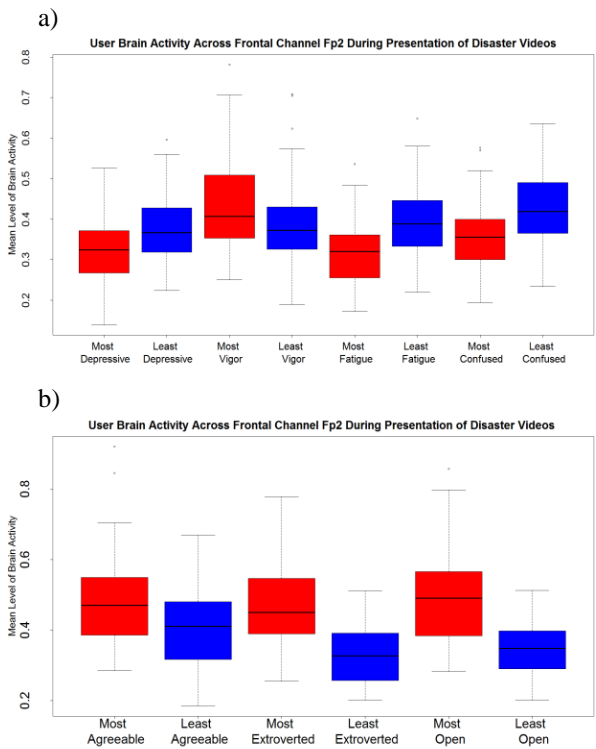

**Fig. 2.** Mean level of RMS brain activity while subjects watch the videos, for users in the lowest (blue) quartile, i.e., least, and highest (red) quartile, i.e., most. All differences are significant to p<0.05. a) for POMS mood traits and b) NEO personality traits

**Table 1.** Total retweet counts for Extroversion and Depression. Extroverts show significantly more willingness to tweet, particularly social tweets. Depressive users show preferences from informative content.

| Tweet Type | Most Extrovert | Least Extrovert | Delta |
|---|---|---|---|
| **Social** | 440 | 206 | 234 |
| **Informative** | 209 | 189 | 20 |
| **Sympathetic** | 143 | 106 | 37 |
| Tweet Type | Most Depressed | Least Depressed | Delta |
| **Social** | 270 | 190 | 80 |
| **Informative** | 242 | 119 | 123 |
| **Sympathetic** | 94 | 91 | 3 |

The present results show that underlying personality and mood may play a significant role in moderating the engagement levels of users.

Table 2. Personality and Mood (Trait) states along with their preferred content (Preference). All noted preferences are statistically significant to at least p < 0.05. The "retweet difference between cohorts" is the difference between the total number retweets made by users in the 1st quartile (i.e., the most) and 4th quartile (i.e., the least) score from each trait. We analyzed content categorized as "social", "informative", or "sympathetic."

| Trait | Preference | Retweet Difference Between Cohorts | | |
|---|---|---|---|---|
| | | Social | Informative | Sympathetic |
| **Agreeable** | More social | 78 | -18 | 20 |
| **Conscientious** | Less content in general, fewer social and informative | -164 | -208 | -97 |
| **Extroversion** | All, particularly social | 234 | 20 | 37 |
| **Anger** | More informative | 27 | 114 | 12 |
| **Confusion** | less content in general, and social content in particular | -104 | -31 | -64 |
| **Depression** | More social and much more informative | 80 | 123 | 3 |
| **Fatigue** | fewer social and sympathetic content, and moderately less informative | -324 | -85 | -139 |
| **Vigor** | All | 96 | 40 | 53 |

We calculated the power spectral density estimation on the subject EEG data to quantify subject attention with the disaster context videos, analyzing the gamma band (30-100 Hz). Gamma waves are strongly associated with intentional attention and cognition, thus by estimating the power density of subjects during each period of video presentation it was possible to quantify different levels of engagement between cohorts of subjects based upon the strength of NEO and POMS personality traits. Subjects in the top quartile for each respective personality trait were compared against their peers in the bottom quartile to determine which group had a greater power density in the gamma wave range, and thus, which group was the most engaged during the videos.

The gamma band had the greatest differences between the cohorts at opposing extremes of the Fatigue and Vigor traits identified by the POMS test. The subjects with the lowest fatigue scores had greater power densities in the gamma wave band across the Fp2, Fp1, F7, and F8 channels, all of which were statistically significant from their peers in the bottom quartile. Likewise, the most vigorous subjects had a statistically significant and greater power density relative to their least vigorous peers. The least depressive subjects only had statistically significant differences in gamma power densities for the F8 and Fp2 channels relative to their most depressive peers (whom had lower power densities). No traits identified by the NEO personality test revealed a statistically significant difference between cohorts, but all traits had overall trends consistent with Fig. 2. For this reason, we do not plot gamma bands, but based on this analysis, we can conclude that certain personality traits indeed confer users with a greater or lesser predisposition to pay attention to content, as shown in Fig. 2.

After the users watch the videos, the testbed presented each user with the relevant messages and alerts in randomized order. Table 1 shows a summary of the relative preference for each type of content. We may safely assume that a user has a "preference" for a specific type of content if they are more likely to retweet that content than other types. Subjects with the most extroverted personalities demonstrated the strongest preference for dismissive messages in the study, as well as the strongest preference for social messages from among the NEO personality types. The most conscientious individuals similarly demonstrated the second strongest preference for social content relative to their least-conscientious peers, and the strongest overall preference for sharing informative posts over social media.

The largest disparities in content preferences, however, were observed for subjects scoring the lowest on the POMS fatigue metric, demonstrating the strongest preference against social and sympathetic posts. The most depressive and angry subjects, conversely, demonstrated the strongest affinity for informative messages relative to their peers scoring the lowest on these dimensions.

Of particular note, the extroversion personality trait and depressive mood showed notable differences between their extremes, shown in Table 2. The most extroverted users were much more likely to retweet any message of a social nature.

Analysis of the EEG data collected during the window of time when subjects were asked if they would retweet a message to their social network reveal that the preferences observed above were often accompanied by significantly different EEG responses as compared to their peers, with the notable exception of depressive personalities, shown Figures 3-8. These figures show the instantaneous power in the F8 channel in users with the lowest and the highest scoring quartiles for each trait on messages they chose to share. The x-axis shows a scaled time such that t=0% is the time of exposure to the message, and t=100% is the moment the user replied. Each user's axis is scaled individually and averaged together with the other users, allowing us to understand engagement over the decision making process.

The levels of relative EEG activity shown in Figs. 3-8 demonstrate that the subject's preferences for certain forms of content exhibited by their responses correspond to particularly high levels of activity over the F8 region of the brain relative to their less responsive peers. As noted earlier, this lone exception to this observation was that the most depressive subjects, which were generally more responsive to informative and social content than their peers, who did not exhibit similarly elevated levels of EEG activity prior to endorsing messages for sharing over their social network. Thus, we see that the notion that users show preferences for sharing messages – and that this preference may be highly sensitive to personality and mood – is corroborated by the users showing increased brain activity *prior to* their decision to retweet for this favored content.

### Related work

The link between social media posting behavior and personality traits has been well established in literature. For example, Big Five personality scores have been used in predicted models based on participant's recent tweets (Golbeck et al. 2011). Similar calculations were run with social graph and interactions between users taken into consideration (Adali and Golbeck 2012). Big Five personality traits were also modeled on abstract groups of users (such as 'listeners', 'popular', 'highly-read' and 'influential') based on user behavior (Quercia et al. 2011). Anti-social traits such as narcissism, psychopathy and Machiavellianism (the "Dark Triad") were predicted and compared with the Big Five personality traits, using language features of tweets (Sumner et al. 2012).

Examination of emotion, personality and brain modeling techniques such as EEG and fMRI has been similarly well established, from predicting patterns of regional brain activity related to extraversion and neuroticism (Schmidtke and Heller 2004), to EEG based emotion recognition when listening to music (Yuan-Pin Lin et al. 2010) or stories designed to evoke specific emotions (Correa et al. 2015; Stikic et al. 2014). Broader emotional recognition with EEG has also been examined with high accuracy (Petrantonakis and Hadjileontiadis 2010; Correa, et al, 2015; Stikic et al, 2014), as well as a functional MRI study of the neuroanatomy of grief (Gündel et al. 2003).

A previous effort at fusing EEG, emotion, and social media focused on producing tweets reflecting a user's emotions at certain physical locations. These tweets included both an emotion component and geotagged location component ("I am Frustrated at this location (Bus Station)") (Almehmadi et al. 2013). Work has been done to tag content based on neurophysiological signals, a technique described in (Yazdani et al. 2009) to produce implicit tagging of emotional states represented in multimedia via EEG and a brain computer interface.

Our present work demonstrates that personality and mood significantly effect that type of content users choose to share under controlled conditions. This shows there is need for models to better characterize user's mood and personality to understand them in live social media feeds. In addition, a broader model of personality and social media would allow us to understand better the friendship paradox (Hodas et al. 2013) and how user-user correlation in personality traits and mood drives social contagion (Kramer et al. 2014).

### Conclusions

Our experiments demonstrate that the personality of the user influences their behavior online in subtle, yet significant, ways. We observe that user's preferences might be predicted from both personality and transitory mood state. This preference is evident in both the brain-activity level (EEG) and in explicit sharing decisions. When constructing an information campaign, the correlation between a user's personality (and mood), interests, and the desired campaign outcome needs to be all taken into account. Because of homophily, most users will be highly correlated with their friends according these very same personality factors. Thus, we will need to understand better the relationship between personality and content preference to better understand and model social behavior online.

It is not surprising that some personalities and moods are more attracted to certain kinds of content. However, this is one of the first results to systematically compare personality measures with content produced during an event – natural disasters, in this case. We also controlled for some of the common confounders that take place during empirical experiments; we were able to account for the correlation between user engagement and preference.

Future work will allow us to further investigate not only the statistical preferences of different users, but also which types of events different personality or moods may be drawn toward when the actively engage with social media. Future modeling may reveal that systematic correlation between the personality of friends may significantly bias local information propagation and information awareness.

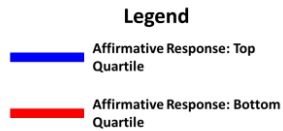

**Legend**
— Affirmative Response: Top Quartile
— Affirmative Response: Bottom Quartile

The following are plots of average squared EEG on the F8 channel for response for messages the users decided to share. We believe F8 the most discriminative channel during retweeting. Each trait is broken down according to the top quartile (users have the "most" of that trait),and bottom quartile (users with the "least" of that trait). *Higher signals indicate more engagement and attention to the message.*

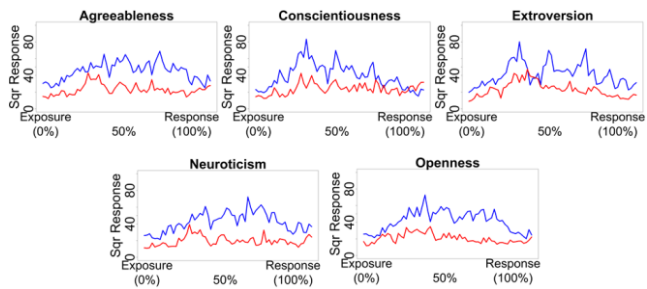

**Fig. 3.** NEO - Informative messages. All time is scaled such that 0% is time of exposure; 100% is time of response.

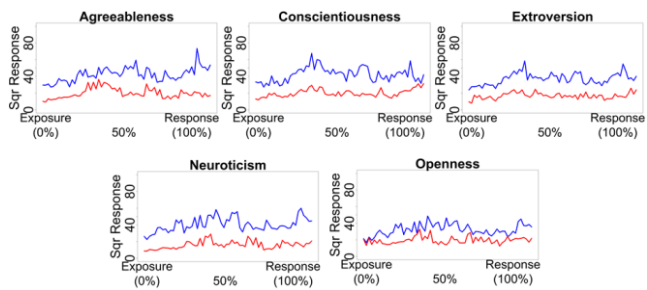

**Fig. 4.** NEO - Social messages.

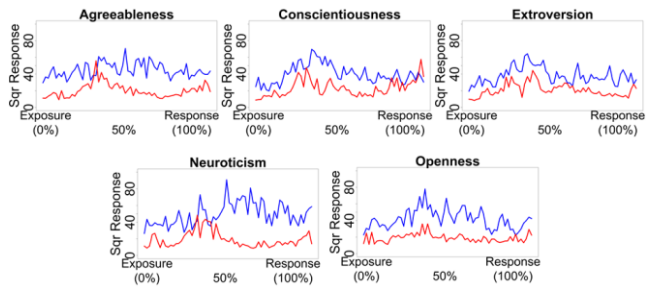

**Fig. 5.** NEO – Sympathetic messages

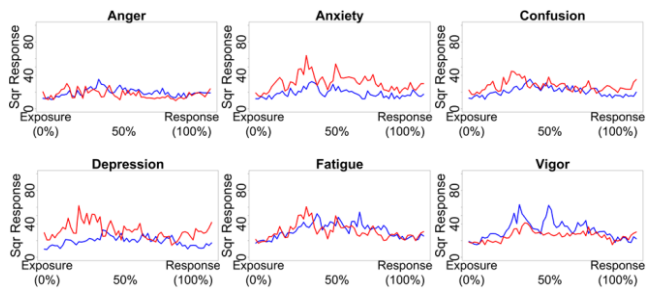

**Fig. 6.** POMS - Informative messages

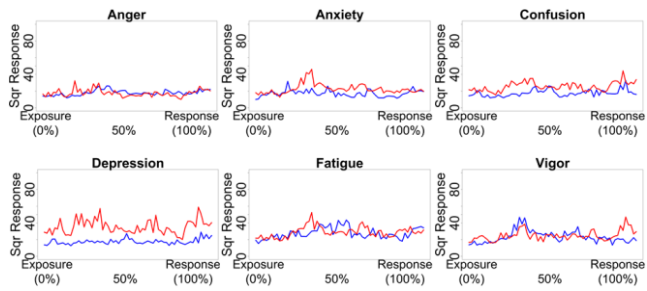

**Fig. 7.** POMS - Social messages

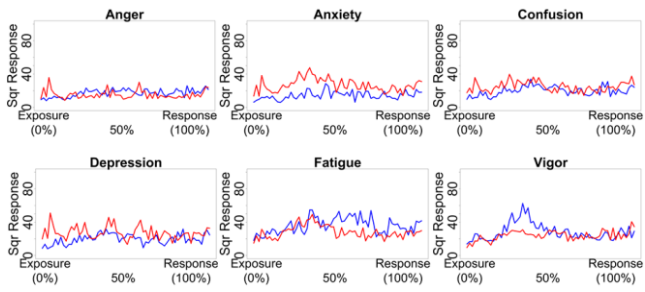

**Fig. 8.** POMS - Sympathetic